\documentclass[twocolumn,showpacs,preprintnumbers,amsmath,amssymb,prb]{revtex4}

\usepackage{graphicx}
\usepackage{dcolumn}
\usepackage{bm}
\usepackage{color}

\begin{document}

\preprint{}

\title{Softening of breathing elastic mode and trigonal elastic mode in disordered pyrochlore magnet NaCaCo$_2$F$_7$}

\author{T. Watanabe$^1$}
\thanks{Electronic address: tadataka@phys.cst.nihon-u.ac.jp}
\author{H. Kato$^1$}
\author{Y. Hara$^2$}
\author{J. W. Krizan$^3$}
\author{R. J. Cava$^3$}
\affiliation{$^1$Department of Physics, College of Science and Technology, Nihon University, Chiyoda, Tokyo 101-8308, Japan}
\affiliation{$^2$National Institute of Technology, Ibaraki College, Hitachinaka 312-8508, Japan}
\affiliation{$^3$Department of Chemistry, Princeton University, Princeton, New Jersey 08544, USA}
\date{\today}

\begin{abstract}
Cobalt pyrochlore fluoride NaCaCo$_2$F$_7$ is a disordered frustrated magnet composed of Co$^{2+}$ ions with an effective spin-$\frac{1}{2}$ magnetic moment and exhibits spin freezing below $T_f \sim$2.4 K. We perform ultrasound velocity measurements on a single crystal of the cubic NaCaCo$_2$F$_7$. The temperature dependence of the bulk modulus (the breathing elastic mode) exhibits Curie-type softening upon cooling below $\sim$20 K down to $T_f$, which is suppressed by the magnetic field. This Curie-type softening should be a precursor to the enhancement of the strength of exchange disorder via the spin-lattice coupling, which causes the spin freezing. In contrast to the magnetic-field-suppressed Curie-type softening in the bulk modulus, the trigonal shear modulus exhibits softening with a characteristic minimum upon cooling, which is enhanced by the magnetic field at temperatures below $\sim$20 K. This magnetic-field-enhanced elastic anomaly in the trigonal shear modulus suggests a coupling of the lattice to the dynamical spin-cluster state. For NaCaCo$_2$F$_7$, the observed elastic anomalies reveal an occurrence of magnetic-field-induced crossover from an isostructural lattice instability toward the spin freezing to a trigonal lattice instability arising from the emergent dynamical spin-cluster state.
\end{abstract}


\maketitle

\section{Introduction}

The antiferromagnetic (AF) pyrochlore lattice of corner-sharing tetrahedra is a canonical geometrically frustrated lattice [\cite{Ramirez}]. The prototypical examples for this lattice system are the spinel oxides with the general formula $AB_2$O$_4$ and the pyrochlore oxides with $A_2B_2$O$_7$, where, whereas only the $B$ site of $AB_2$O$_4$ forms the pyrochlore lattice, both the $A$ and $B$ sites of $A_2B_2$O$_7$ independently form the pyrochlore lattice. Many spinel and pyrochlore oxides studies have respectively been devoted to the 3$d$ transition-metal and the 4$f$ rare-earth oxides, where the intersite magnetic interaction in the 3$d$ transition-metal oxide is stronger than that in the 4$f$ rare-earth oxide [\cite{Lee0,Gardner}].

The recently discovered pyrochlore fluorides NaSr$B_2$F$_7$ ($B$ = Mn, Fe, and Co) and NaCa$B_2$F$_7$ ($B$ = Fe, Co, and Ni) provide a new platform for the study of 3$d$ transition-metal frustrated magnets, where the large single crystals are available for experimental studies [\cite{Krizan1,Krizan2,Krizan3,Sanders}]. In this family, despite the AF Weiss temperature $\theta_W\sim -140$ K (NaCaCo$_2$F$_7$) to $-73$ K (NaCaFe$_2$F$_7$), the spin freezing occurs at low temperatures below $T_f\sim$ 2.4 K (NaCaCo$_2$F$_7$) to 3.9 K (NaCaFe$_2$F$_7$), indicating the presence of strong frustration with $|\theta_W/T_f|\sim$ 19 to 58 [\cite{Krizan1,Krizan2,Krizan3,Sanders}]. For NaSr$B_2$F$_7$ and NaCa$B_2$F$_7$, whereas the magnetic $B^{2+}$ ions uniformly occupy the pyrochlore $B$-sites of the $A_2B_2$F$_7$ structure, the nonmagnetic Na$^+$ and Sr$^{2+}$/Ca$^{2+}$ ions are randomly distributed on the pyrochlore $A$ sites. Thus, the magnetism of NaSr$B_2$F$_7$ and NaCa$B_2$F$_7$ should be influenced by the inherently present exchange disorder, which can lead to spin freezing [\cite{Ratcliff,Saunders,Andreanov,Shinaoka,Silverstein}].

The cobalt pyrochlore fluoride NaCaCo$_2$F$_7$ is a disordered frustrated magnet with $\theta_W\sim-140$ K and $T_f\sim 2.4$ K, which has the largest $|\theta_W/T_f|\sim$ 58 among the pyrochlore fluorides NaSr$B_2$F$_7$ and NaCa$B_2$F$_7$ [\cite{Krizan1}]. For this compound, inelastic neutron scattering (INS) experiments revealed that the single ion ground state of Co$^{2+}$ is a Kramers doublet with an XY-like effective spin-$\frac{1}{2}$ magnetic moment, which is generated by spin-orbit coupling [\cite{Ross2}]. Furthermore, the INS experiments in NaCaCo$_2$F$_7$ revealed the formation of static short-range AF clusters with XY character below $T_f$, and the persistence of a dynamical spin-cluster state above $T_f$ [\cite{Ross1,Frandsen}]. The presence of strong dynamic correlations above $T_f$ in NaCaCo$_2$F$_7$ was also evidenced by NMR and ESR studies [\cite{Sarkar,Zeisner}]. Additionally, above $T_f$, the ESR experiments suggested the coexistence of a gapless excitation mode of a cooperative paramagnetic state and a low-energy gapped excitation mode in the order of sub meV, arising from the strong frustration [\cite{Zeisner}].

In this paper, we present ultrasound velocity measurements of the cobalt pyrochlore fluoride NaCaCo$_2$F$_7$, where we determine the elastic moduli of this compound. The sound velocity or the elastic modulus is a useful probe enabling symmetry-resolved thermodynamic information to be extracted from a crystal [\cite{Luthi}]. Furthermore, as the ultrasound velocity can be measured with a high precision of $\sim$ppm, its measurements can sensitively probe elastic anomalies driven by phase transition, fluctuations, and excitations [\cite{Luthi}]. For the frustrated magnets, the ultrasound velocity measurements have proven to be a useful tool for studying not only the ground state but also the excited states [\cite{Nii,Watanabe1,Watanabe2,Watanabe3,Watanabe4,Watanabe5,Watanabe6,Watanabe7}]. In the present study on NaCaCo$_2$F$_7$, we find two different types of elastic anomalies at low temperatures above $T_f$, which reveal, respectively, an isostructural lattice instability toward the spin freezing at $T_f$ and a trigonal lattice instability arising from the emergent dynamical spin-cluster state. Additionally, the magnetic field dependence of the observed elastic anomalies reveals an occurrence of magnetic-field-induced crossover of the dominant lattice instability from isostructural to trigonal.

\section{Experimental}

Single crystals of NaCaCo$_2$F$_7$ were prepared by the optical floating-zone method [\cite{Krizan1}]. The ultrasound velocities were measured utilizing the phase-comparison technique, where the ultrasound velocity or the elastic modulus can be measured with a high precision of $\sim$ppm. For the measurements, the longitudinal and transverse ultrasounds at a frequency of 30 MHz were generated and detected by LiNbO$_3$ transducers attached on the parallel mirror surfaces of the single-crystalline sample. We measured the ultrasound velocities in all the symmetrically independent elastic moduli in the cubic crystal, specifically, compression modulus $C_{11}$, tetragonal shear modulus $\frac{C_{11}-C_{12}}{2} \equiv C_t$, and trigonal shear modulus $C_{44}$. From $C_{11}$ and $C_t$ data, we also obtained the bulk modulus $C_B = \frac{C_{11}+2C_{12}}{3}=C_{11}-\frac{4}{3}C_t$. The respective measurements of $C_{11}$, $C_t$, and $C_{44}$ were performed using longitudinal ultrasound with propagation {\bf k}$\parallel$[100] and polarization {\bf u}$\parallel$[100], transverse ultrasound with {\bf k}$\parallel$[110] and {\bf u}$\parallel$[1$\bar{1}$0], and transverse ultrasound with {\bf k}$\parallel$[110] and {\bf u}$\parallel$[001]. The sound velocities of NaCaCo$_2$F$_7$ measured at room temperature (300 K) are $\sim$5640 m/s for $C_{11}$, $\sim$3010 m/s for $C_t$, and $\sim$2950 m/s for $C_{44}$.

\section{Results and Discussion}

Figures 1(a)--1(c) respectively present the temperature ($T$) dependence of the elastic moduli $C_B(T)$, $C_t(T)$, and $C_{44}(T)$ with zero magnetic field ($H$ = 0) in NaCaCo$_2$F$_7$. Here $C_B(T)=C_{11}(T)-\frac{4}{3}C_t(T)$ is obtained from $C_{11}(T)$ [Fig. 5(a)] (Appendix) and $C_t(T)$ [Fig. 1(b)]. In Figs. 1(a)--1(c), all the elastic moduli exhibit monotonic hardening upon cooling down to about 10$-$20 K, as is usually observed in solids [\cite{Varshni}]. However, at low temperatures below about 10$-$20 K, the elastic moduli exhibit elastic-mode-dependent unusual softening upon cooling [the insets in Figs. 1(a)--1(c)]. Fig. 1(d) compares the softening magnitudes in $C_B(T)$, $C_t(T)$, and $C_{44}(T)$ with $H$ = 0. Here it is evident that the softening magnitude in $C_B(T)$ ($\frac{\Delta C_B}{C_B}\sim 4300$ ppm) is much larger than that in $C_t(T)$ ($\frac{\Delta C_t}{C_t}\sim 800$ ppm) and $C_{44}(T)$ ($\frac{\Delta C_{44}}{C_{44}}\sim 120$ ppm).

Figures 2(a) and 2(b) respectively depict $C_B(T)$ and $C_{t}(T)$ with $H||$[110] below 20 K in NaCaCo$_2$F$_7$. Here $C_B(T)=C_{11}(T)-\frac{4}{3}C_t(T)$ is obtained from $C_{11}(T)$ [Fig. 5(b)] (Appendix) and $C_t(T)$ [Fig. 2(b)]. At $H$ = 0, $C_B(T)$ and $C_{t}(T)$ exhibit softening upon cooling below $\sim$20 K and $\sim$15 K, respectively, but turn to hardening below $\sim$2.4 K. This softening-to-hardening turning temperature corresponds to the spin-freezing temperature $T_f\sim$ 2.4 K determined from the dc/ac magnetic susceptibility and specific heat measurements, as indicated in Figs. 2(a) and 2(b) [\cite{Krizan1}]. Thus, the softening-to-hardening turning at $\sim$2.4 K in $C_B(T)$ and $C_{t}(T)$ with $H$ = 0 should be a result of the spin freezing, and the softening above $\sim$2.4 K should be its precursor. The application of $H$, as shown in Figs. 2(a) and 2(b) with the dotted arrows, suppresses the softening as well as the softening-to-hardening turning in $C_B(T)$ and $C_{t}(T)$, which should arise from the suppression of the spin-freezing behavior by $H$. At temperatures above $\sim$20 K, $C_B(T)$ and $C_{t}(T)$ are independent of magnetic field (not shown).

\begin{figure}[t]
\begin{center}
\includegraphics[scale=0.39]{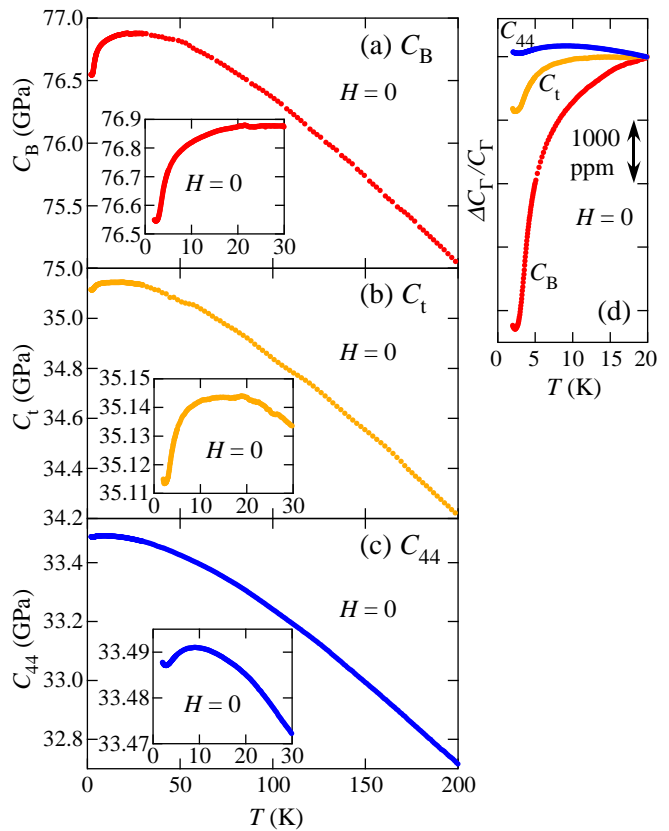}
\caption{\label{fig:fig3} (Color online) (a)--(c) Elastic moduli of NaCaCo$_2$F$_7$ as functions of $T$ with $H = 0$. (a) $C_B(T)$, (b) $C_t(T)$, and (c) $C_{44}(T)$. The insets in (a)--(c) respectively depict the expanded views of $C_B(T)$, $C_t(T)$, and $C_{44}(T)$ with $H = 0$ below 30 K. (d) Comparison of the relative shifts of $C_B(T)$, $C_t(T)$, and $C_{44}(T)$ with $H = 0$ below 20 K.}
\end{center}
\end{figure}

For NaCaCo$_2$F$_7$, taking into consideration that the magnitude of the softening at $H$ = 0 in $C_B(T)$ is much larger than that in $C_t(T)$ and $C_{44}(T)$ [Fig. 1(d)], the precursor softening to the spin freezing above $T_f$ should be characterized as a softening in the bulk modulus $C_B(T)$. As seen in Fig. 2(a), the softening in $C_B(T)$ with $H$ = 0 behaves as $C_B(T)\sim -1/T$. In magnets, such a Curie-type softening emerges as a precursor to a structural transition, which is driven by the coupling of the lattice to the electronic degrees of freedom [\cite{Luthi,Kino,Kataoka,Hazama,Nii,Watanabe3,Watanabe4,Watanabe7}]. A mean-field expression of the Curie-type softening in the temperature dependence of the elastic modulus $C_{\Gamma}(T)$ is given as
\begin{equation}
C_{\Gamma}(T) = C_{\Gamma}^{(0)} \frac{T-T_c}{T-\theta},
\label{eq:Curie}
\end{equation}
where $C_{\Gamma}^{(0)}$ is the background elastic constant, $T_c$ is the second-order critical temperature for elastic softening $C_{\Gamma}\rightarrow$ 0, and $\theta$ is the intersite strain-sensitive magnetic interaction. In Fig. 2(a), a fit of the experimental $C_B(T)$ with $H$ = 0 to Eq. (1) for $T > T_f$ is drawn as a solid black curve, which, with the fit parameter values listed in this figure, excellently reproduces the experimental data.

\begin{figure}[t]
\begin{center}
\includegraphics[scale=0.4]{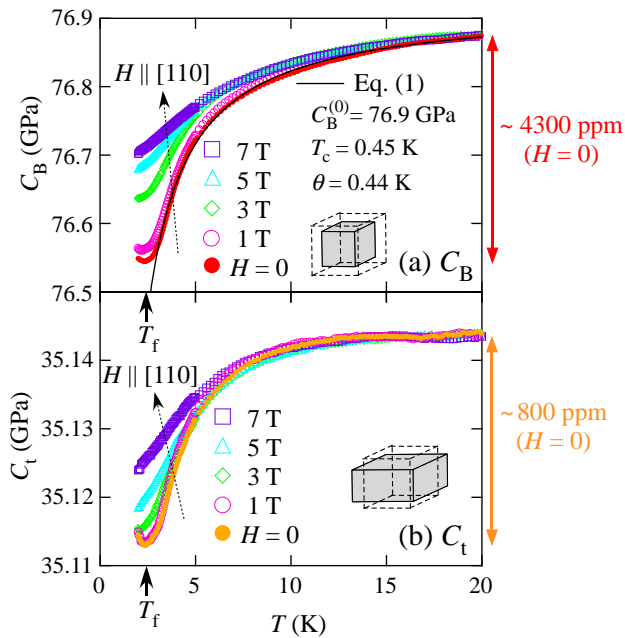}
\caption{\label{fig:fig3} (Color online) (a) $C_B(T)$ and (b) $C_t(T)$ of NaCaCo$_2$F$_7$ with $H||$[110] below 20 K. The dotted arrows in (a) and (b) are guides to the eye, indicating the variations of $C_B(T)$ and $C_t(T)$ with increasing $H$. $T_f$ in (a) and (b) indicates the spin-freezing temperature determined from the dc/ac magnetic susceptibility and specific heat measurements [\cite{Krizan1}]. The solid curve in (a) is a fit of the zero-field experimental $C_B(T)$ to Eq. (1) below 20 K. The values of the fit parameters are also listed in (a). The inset pictures in (a) and (b) respectively illustrate schematics of the volume strain in $C_B$ and the tetragonal strain in $C_t$. The solid double arrows at the right side of (a) and (b) respectively indicate the softening magnitudes in the zero-field $C_B(T)$ and $C_t(T)$.}
\end{center}
\end{figure}

For NaCaCo$_2$F$_7$, the INS experiments confirmed that the single-ion ground state of Co$^{2+}$ is a Kramers doublet with the effective spin-$\frac{1}{2}$ magnetic moment, which is generated by spin-orbit coupling [\cite{Ross2}]. Thus, one possible origin for the Curie-type softening in NaCaCo$_2$F$_7$ is the Jahn--Teller effect, which is driven by the coupling of the lattice to the degenerate single-ion state (the quadrupole-lattice coupling) [\cite{Luthi}]. In this scenario, the Curie-type softening is a precursor to the lowering of the lattice symmetry, which lifts the degeneracy of the single-ion ground state, and this precursor softening should occur in the symmetry-lowering elastic mode such as, in the cubic lattice, $C_t(T)$ and $C_{44}(T)$. Thus, as a possible origin for the Curie-type softening in {\it the symmetry-conserving ``breathing'' elastic mode} $C_B(T)$ of NaCaCo$_2$F$_7$, the Jahn--Teller effect is ruled out.

Consequently, the Curie-type softening in NaCaCo$_2$F$_7$ is most probably explained by assuming a coupling of ultrasound with the magnetic ions through the magnetoelastic coupling acting on the exchange interactions, where the exchange striction arises from an ultrasound modulation of the exchange interactions [\cite{Luthi,Watanabe3,Watanabe5}]. That is, the softening in NaCaCo$_2$F$_7$ should be driven by the pseudospin-lattice coupling. One possible origin for such a softening is the so-called spin Jahn--Teller effect [\cite{Watanabe3,Yamashita,Tchernyshyov,Ji}]. In this scenario, the Curie-type softening is a precursor to the magnetostructural transition, where the spin-lattice coupling lowers the crystal symmetry resulting in the release of frustration [\cite{Watanabe3}]. Additionally, this precursor softening should occur in the symmetry-lowering elastic mode but not in the breathing elastic mode $C_B(T)$, which is similar to the quadrupolar Jahn--Teller effect mentioned in the preceding paragraph. Thus, the Curie-type softening in the breathing elastic mode $C_B(T)$ of NaCaCo$_2$F$_7$ is uniquely different from the Curie-type softening originated from the spin Jahn--Teller effect.

For NaCaCo$_2$F$_7$, the observation of the Curie-type softening in the breathing elastic mode $C_B(T)$ indicates the presence of an isostructural lattice instability, which is a precursor to the spin freezing. Taking into consideration the inherent presence of exchange disorder in NaCaCo$_2$F$_7$ due to the random occupation of Na$^+$ and Ca$^{2+}$ on the pyrochlore $A$ sites of the $A_2B_2$F$_7$ structure, the most natural explanation for the isostructural lattice instability is {\it a precursor to the enhancement of the strength of exchange disorder via the spin-lattice coupling, which causes the spin freezing.} Such a unique magnetoelastic effect in the disordered frustrated magnet is expected to occur in not only NaCaCo$_2$F$_7$ but also other families of the disordered pyrochlore fluorides, namely, NaSr$B_2$F$_7$ ($B$ = Mn, Fe, and Co) and NaCa$B_2$F$_7$ ($B$ = Fe and Ni), which all exhibit spin freezing at a temperature $T_f$ much lower than the Weiss temperature $|\theta_W|$, $|\theta_W/T_f|\sim$ 19 to 58 [\cite{Krizan1,Krizan2,Krizan3,Sanders}].

In addition to the above-mentioned Curie-type softening in $C_B(T)$, for NaCaCo$_2$F$_7$, we also find another kind of intriguing elastic anomaly in the trigonal shear modulus $C_{44}(T)$. At zero magnetic field, as already mentioned in conjunction with Fig. 1(d), the magnitude of the softening in $C_{44}(T)$ is much smaller than that in $C_B(T)$. However, the application of $H$ enormously enhances the softening in $C_{44}(T)$. Figure 3 depicts $C_{44}(T)$ with $H||$[110] below 20 K. As shown in Fig. 3(a), the softening in $C_{44}(T)$ is enhanced by $H$ below $\sim$20 K, which is opposite to the suppression of the softening by $H$ in $C_B(T)$ and $C_t(T)$ [Figs. 2(a) and 2(b)]. The softening magnitude in the 7 T $C_{44}(T)$ ($\sim$1600 ppm) is about 13 times larger than that in the zero-field $C_{44}(T)$ ($\sim$120 ppm) [Figs. 3(a) and 3(b)]. This $H$-enhanced softening in $C_{44}(T)$ should have an origin different from the precursor to the spin freezing that is the origin of the $H$-suppressed softening in $C_B(T)$. At temperatures above $\sim$20 K, $C_{44}(T)$ is independent of magnetic field (not shown).

\begin{figure}[t]
\begin{center}
\includegraphics[scale=0.4]{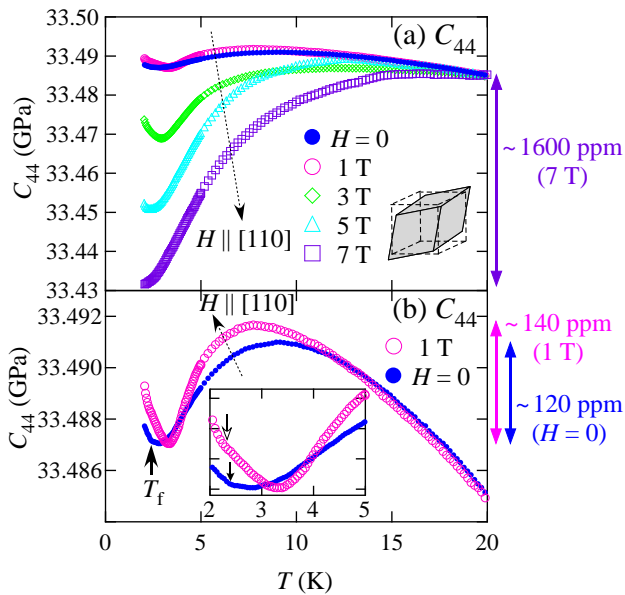}
\caption{\label{fig:fig3} (Color online) $C_{44}(T)$ of NaCaCo$_2$F$_7$ with $H||$[110] below 20 K. (a) $H=0\sim$ 7 T and (b) $H=0$ and 1 T. The dotted arrows in (a) and (b) are guides to the eye, indicating the variation of $C_{44}(T)$ with increasing $H$. The inset picture in (a) illustrates a schematic of the trigonal strain in $C_{44}$. $T_f$ in (b) indicates the spin-freezing temperature determined from the dc/ac magnetic susceptibility and specific heat measurements [\cite{Krizan1}]. The solid double arrows at the right side of (a) and (b) respectively indicate the softening magnitudes in the 7 T, 1 T, and zero-field $C_{44}(T)$. The inset in (b) depicts the expanded view of the zero-field and 1 T $C_{44}(T)$ in 2 K $<T<$ 5 K, where the solid and open arrows respectively indicate the slope-change temperatures in the zero-field and 1 T $C_{44}(T)$.}
\end{center}
\end{figure}

Here, we look more closely at the observed $H$ variation of the softening in $C_{44}(T)$. Figure 3(b) presents the expanded view of $C_{44}(T)$ with $H=0$ and 1 T below 20 K. In Fig. 3(b), whereas $C_{44}(T)$ with $H=0$ and that with 1 T identically harden upon cooling down to $\sim$12 K, the starting temperature of the softening in the 1 T $C_{44}(T)$ is lowered compared to that in the zero-field $C_{44}(T)$, which indicates the suppression of the softening by $H$ [the dotted arrow in Fig. 3(b)]. This $H$-suppressed softening component should correspond to a precursor to the spin freezing, namely, the Curie-type softening, which is similar to the softening in $C_B(T)$ and $C_t(T)$ [Figs. 2(a) and 2(b)]. However, in Fig. 3(b), the softening in the 1 T $C_{44}(T)$ below $\sim$7 K is steeper than that in the zero-field $C_{44}(T)$, and the softening magnitude in the 1 T $C_{44}(T)$ ($\sim$140 ppm) is larger than that in the zero-field $C_{44}(T)$ ($\sim$120 ppm), which indicates the presence of the $H$-enhanced softening component. Thus, the comparison of the zero-field and 1 T $C_{44}(T)$ in Fig. 3(b) indicates the presence of not only $H$-enhanced but also $H$-suppressed softening components in $C_{44}(T)$, and the $H$-enhanced component should become dominant at higher $H$, as is seen in the 3 T, 5 T, and 7 T $C_{44}(T)$ in Fig. 3(a).

As seen in Fig. 3(b), $C_{44}(T)$ with $H=0$ and 1 T exhibits its minimum at $\sim$3 K. The inset in Fig. 3(b) presents the expanded view of the zero-field and 1 T $C_{44}(T)$ in 2 K $<T<$ 5 K. The minimum point temperature in $C_{44}(T)$ with $H=0$ and that with 1 T are $\sim$2.9 K and $\sim$3.3 K, respectively, which are a little higher than the spin-freezing temperature $T_f\sim$ 2.4 K [Fig. 3(b)]. Thus, the minimum in $C_{44}(T)$ should be caused by an origin other than the spin freezing. At $\sim$2.4 K, $C_{44}(T)$ with $H=0$ and 1 T exhibits a small slope change [the solid and open arrows in the inset in Fig. 3(b)], which should arise from the spin freezing. Consequently, the $H$-enhanced softening component in $C_{44}(T)$ with $H=0$ and 1 T should be characterized as a nonmonotonic softening, which exhibits the elasticity minimum at around $\sim$3 K. Additionally, we can see in Fig. 3(a) that this $H$-enhanced component becomes dominant in the 3 T, 5 T, and 7 T $C_{44}(T)$, where the minimum point temperature is lowered with increasing $H$.

Similar to the $H$-suppressed Curie-type softening in $C_B(T)$, the $H$-enhanced softening with minimum elastic anomaly in $C_{44}(T)$ should also arise via the exchange striction mechanism, but it should have an origin other than the precursor to the spin freezing. The most probable origin for the softening with minimum in $C_{44}(T)$ is the coupling between the correlated paramagnetic state and the acoustic phonons. For NaCaCo$_2$F$_7$, the INS [\cite{Ross1}], NMR [\cite{Sarkar}], and ESR [\cite{Zeisner}] studies revealed the presence of short-range dynamical magnetic correlations above $T_f\sim$2.4 K. Additionally, the INS study also revealed that this correlated paramagnetic state consists of AF XY spin clusters, which become static below $T_f$ [\cite{Ross1,Ross2}]. Thus, the softening with minimum in $C_{44}(T)$ should originate from {\it the coupling of the lattice to the dynamical short-range XY clusters.}

It is noted that, similar to NaCaCo$_2$F$_7$, the softening with minimum in $C_{\Gamma}(T)$ is also observed in the frustrated spinel oxides, the origin of which is considered to be the coupling of the lattice to the spin-cluster excitations via the exchange striction mechanism [\cite{Watanabe2,Watanabe3,Watanabe4,Watanabe5,Watanabe6}]. This spin-cluster-driven elastic softening is generally explained as the presence of a finite gap for the excitations, which is sensitive to strain [\cite{Watanabe3}]. In the mean-field approximation, $C_{\Gamma}(T)$ in the spin-cluster system is written as [\cite{Watanabe3}]
\begin{equation}
C_{\Gamma}(T)=C_{\Gamma}^{(0)}-G_{\Gamma}^2N\frac{\chi_{\Gamma}(T)}{[1-K_{\Gamma}\chi_{\Gamma}(T)]},
\label{eq:SM}
\end{equation}
where $C_{\Gamma}^{(0)}$ is the background elastic constant, $N$ is the density of spin clusters, $G_{\Gamma}=|\partial \Delta/\partial \epsilon_{\Gamma}|$ is the coupling constant for a single spin cluster measuring the strain ($\epsilon_{\Gamma}$) dependence of the excitation gap $\Delta$, $K_{\Gamma}$ is the inter-spin-cluster interaction, and $\chi_{\Gamma}(T)$ is the strain susceptibility of a single spin cluster. From Eq.~(\ref{eq:SM}), when $C_{\Gamma}(T)$ strongly couples to the excited state at $\Delta$, this elastic mode exhibits softening upon cooling roughly down to $T\sim\Delta$ but recovery of the elasticity (hardening) roughly below $T\sim\Delta$; $C_{\Gamma}(T)$ exhibits a minimum roughly at $T\sim\Delta$.

We now analyze the softening with minimum in $C_{44}(T)$ in NaCaCo$_2$F$_7$ using Eq. (2). In the frustrated spinel oxides, the observation of the spin-cluster excitations by the INS experiments revealed that the number of magnetic ions, shape, and symmetry of spin clusters vary from compound to compound depending on the dominant exchange path [\cite{Lee1,Tomiyasu1,Tomiyasu2,Tomiyasu3,Tomiyasu4,Tomiyasu5,Tomiyasu6}]. For NaCaCo$_2$F$_7$, the INS study revealed that the correlation length of the dynamical short-range XY clusters in the paramagnetic phase ($\sim$8 $\AA$) is larger than the length scale of the single Co$^{2+}$ tetrahedron ($\sim$3.5 $\AA$) [\cite{Ross1}]. Additionally, the $Q$-space INS pattern in the paramagnetic phase of NaCaCo$_2$F$_7$ looks like that of the AF seven-spin clusters (spin heptamers) proposed in the chromite spinel MgCr$_2$O$_4$ with the isomorphic magnetic pyrochlore lattice as the Co$^{2+}$ sites in NaCaCo$_2$F$_7$ [\cite{Ross1,Ross2,Tomiyasu1,Tomiyasu2}]. Thus, although the exact shape of the spin cluster in NaCaCo$_2$F$_7$ has not yet been identified, we assume the AF spin heptamer for the analysis, which consists of two corner-sharing Co$^{2+}$ tetrahedra [the inset picture in Fig. 4(a)]. We note here that, from the symmetry point of view, the spin-heptamer excitations should couple more sensitively to the trigonal lattice deformations, which is compatible with the selective observation of the softening with minimum in the trigonal shear modulus $C_{44}(T)$ in the present study.

\begin{figure}[t]
\begin{center}
\includegraphics[scale=0.4]{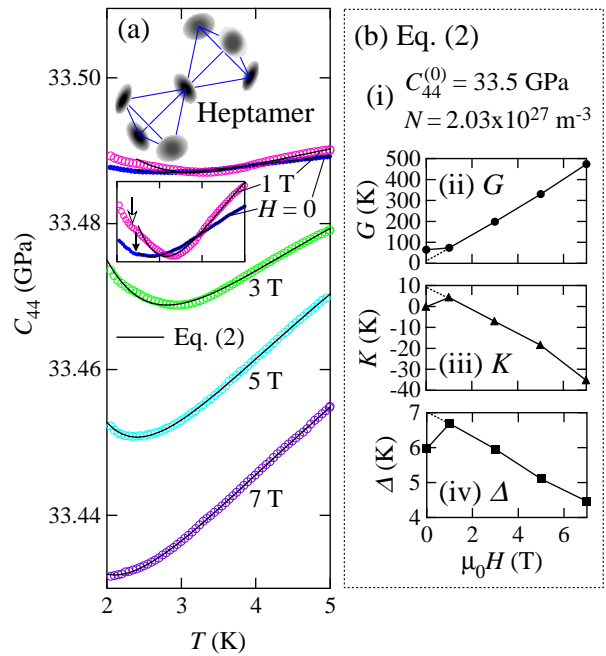}
\caption{\label{fig:fig4} (Color online) (a) $C_{44}(T)$ of NaCaCo$_2$F$_7$ with $H||$[110] in 2 K $<T<$ 5 K [circles, from Fig. 3(a)]. The solid curves are fits to Eq. (2) with the assumption of singlet--triplet gapped excitations of AF spin heptamers. The inset picture illustrates the single XY heptamer. The inset figure depicts the expanded view of the zero-field and 1 T $C_{44}(T)$, where the solid and open arrows respectively indicate the slope-change temperatures in the zero-field and 1 T $C_{44}(T)$. (b) Fit parameter values for the fit curves in (a). (i) $C_{44}^{(0)}$ and $N$, (ii) $G$, (iii) $K$, and (iv) $\Delta$ in Eq. (2). The values of $G$, $K$, and $\Delta$ are respectively plotted in (ii), (iii), and (iv) as functions of $H$. The dotted lines in (ii)--(iv) indicate linear extrapolations of the plots above $\mu_0 H$ = 1 T to $H$ = 0.}
\end{center}
\end{figure}

Figure 4(a) depicts the experimental data of $C_{44}(T)$ with $H||$[110] in 2 K $<T<$ 5 K [circles, from Fig. 3(a)] and their fits to Eq. (2) assuming spin heptamer excitations with a singlet--triplet excitation gap $\Delta$ [solid curves]. The parameter values for the fit curves in Fig. 4(a) are shown in Fig. 4(b). For NaCaCo$_2$F$_7$, the density of the spin heptamers is assumed to be $N=2.03\times10^{27}$ m$^{-3}$ [Fig. 4(b)(i)], which is one-seventh of the density of Co$^{2+}$ ions in NaCaCo$_2$F$_7$. The fittings are performed at temperatures below 5 K because, among the experimental data of $C_{44}(T)$ with $H||$[110] [Figs. 3(a) and 3(b)], the lowest starting temperature of the softening is $\sim$7 K for the 1 T $C_{44}(T)$ [Fig. 3(b)]. Whereas the lower limit of the temperature range of the fitting is 2 K for the 3 T, 5 T, and 7 T $C_{44}(T)$, that for the zero-field and 1 T $C_{44}(T)$ is the slope-change temperature of $T_f\sim$ 2.4 K [the solid and open arrows in the inset figure in Fig. 4(a)]. As clearly seen in Fig. 4(a), the fits of Eq. (2) are in excellent agreement with the experimental data, reproducing the softening with minimum in $C_{44}(T)$.

Figures 4(b)(ii)--(iv) respectively display plots of the fit values of $G$, $K$, and $\Delta$ in Eq. (2) for the fit curves in Fig. 4(a) as functions of $H$. These plots in Figs. 4(b)(ii)--(iv) respectively exhibit monotonic $H$ variations at $\mu_0H$ = 1 T $\sim$ 7 T, but the values at $H$ = 0 deviate from the respective monotonic $H$ variations. These deviations at $H$ = 0 are probably due to the presence of the Curie-type softening in the zero-field $C_{44}(T)$ in addition to the softening with minimum, as was mentioned in conjunction with Fig. 3(b). The ``correct'' values of $G$, $K$, and $\Delta$ at $H$ = 0 are expected to follow the monotonic $H$ variations at $\mu_0H$ = 1 T $\sim$ 7 T, which are indicated in Figs. 4(b)(ii)--(iv) as dotted lines between $H$ = 0 and 1 T.

As seen in Fig. 4(b)(ii), the intra-heptamer coupling $G$ is enhanced with increasing $H$, which corresponds to the enhancement of the softening magnitude in $C_{44}(T)$ by $H$ [Fig. 3(a)]. Accordingly, as seen in Fig. 4(b)(iii), the application of $H$ also enhances the inter-heptamer antiferrodistortive interaction $K$ ($<0$). These $H$ variations of $G$ and $K$ indicate that the application of $H$ enhances the trigonal lattice instability, which is driven by the coupling of the lattice to the dynamical spin-cluster state. Here we note again that, in contrast to the $H$-enhanced trigonal lattice instability [Fig. 3(a)], the application of $H$ suppresses the spin freezing and its precursor of the isostructural lattice instability, namely, the Curie-type softening in the breathing elastic mode $C_B(T)$ [Fig. 2(a)]. For NaCaCo$_2$F$_7$, the present study reveals {\it an occurrence of $H$-induced crossover from an isostructural lattice instability toward the spin freezing to a trigonal lattice instability arising from the emergent dynamical spin-cluster state.}

As seen in Fig. 4(b)(iv), the excitation gap of the single heptamer $\Delta$ decreases with increasing $H$. From this $H$ variation, assuming the $H$-linear decrease of $\Delta$, it is expected that a continuous transition from the gapped ground state to the gapless state occurs at $\sim$19 T. On the other hand, assuming the isolated spin heptamers, it is expected from the gap value of $\Delta\sim$7 K at $H$ = 0 [the dotted extrapolation line in Fig. 4(b)(iv)] that the transition from the gapped ground state to the gapless state occurs at $\sim$5 T. Thus, the $H$ dependence of $\Delta$ shown in Fig. 4(b)(iv) indicates the presence of inter-heptamer interaction, which is compatible with the enhancement of the magnitude of $K$ with increasing $H$ [Fig. 4(b)(iii)].

The entire family of the disordered pyrochlore fluorides NaSr$B_2$F$_7$ ($B$ = Mn, Fe, and Co) and NaCa$B_2$F$_7$ ($B$ = Fe, Co, and Ni) exhibits spin freezing, which should be a result of the inherent presence of exchange disorder [\cite{Krizan1,Krizan2,Krizan3,Sanders}]. However, as the single-ion state of the pyrochlore $B^{2+}$ site varies from compound to compound, the correlated magnetic state of NaSr$B_2$F$_7$ and NaCa$B_2$F$_7$ should vary from compound to compound. For instance, unlike the spin-$\frac{1}{2}$ XY magnet NaCaCo$_2$F$_7$, the nickel pyrochlore fluoride NaCaNi$_2$F$_7$ is considered to be a spin-1 Heisenberg magnet and is suggested to be a three-dimensional quantum spin liquid candidate [\cite{Plumb,Zhang}]. Thus, for NaSr$B_2$F$_7$ and NaCa$B_2$F$_7$, the bulk modulus $C_B(T)$ of the entire family is expected to exhibit the Curie-type of precursor softening to spin freezing, but the elastic anomaly driven by the correlated magnetic state is expected to vary from compound to compound, which in the case of NaCaCo$_2$F$_7$ is the softening with minimum in $C_{44}(T)$.

\section{Summary}

Ultrasound velocity measurements of NaCaCo$_2$F$_7$ revealed elastic anomalies in the bulk modulus $C_B(T)$ and the trigonal shear modulus $C_{44}(T)$, which are respectively suppressed and enhanced by the magnetic field at temperatures below $\sim$20 K. These anomalies indicated the occurrence of the magnetic-field-induced crossover from the isostructural to the trigonal lattice instability. The isostructural lattice instability indicated that the spin freezing in NaCaCo$_2$F$_7$ is driven by the enhancement of the strength of exchange disorder via the spin-lattice coupling, which is unique to the disordered frustrated magnet. The trigonal lattice instability suggested the coupling of the lattice to the dynamical spin-cluster state. It is inferred from the present study that the isostructural lattice instability emerges in the entire family of disordered pyrochlore fluorides NaSr$B_2$F$_7$ ($B$ = Mn, Fe, and Co) and NaCa$B_2$F$_7$ ($B$ = Fe, Co, and Ni), which gives rise to the spin freezing. Furthermore, taking into consideration that the single-ion state of the pyrochlore $B^{2+}$ site in NaSr$B_2$F$_7$ and NaCa$B_2$F$_7$ varies from compound to compound, the correlated magnetic state of NaSr$B_2$F$_7$ and NaCa$B_2$F$_7$ should vary from compound to compound, which is expected to give rise to compound-dependent lattice instabilities.

\begin{figure}[b]
\begin{center}
\includegraphics[scale=0.4]{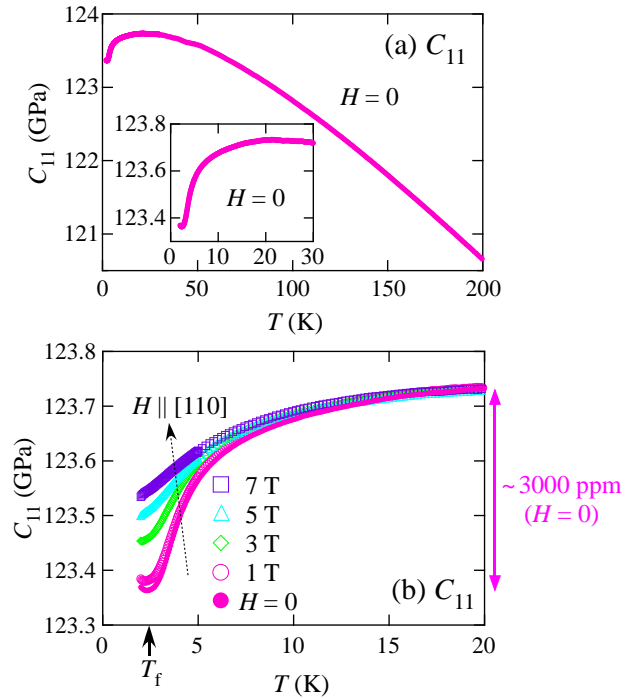}
\caption{\label{fig:fig5} (Color online) $C_{11}(T)$ of NaCaCo$_2$F$_7$ with (a) $H=0$ below 200 K and (b) $H||$[110] below 20 K. The inset in (a) depicts the expanded view of $C_{11}(T)$ with $H=0$ below 30 K. The dotted arrow in (b) is a guide to the eye, indicating the variation of $C_{11}(T)$ with increasing $H$. $T_f$ in (b) indicates the spin-freezing temperature determined from the dc/ac magnetic susceptibility and specific heat measurements [\cite{Krizan1}]. The solid double arrow at the right side of (b) indicates the softening magnitude in the zero-field $C_{11}(T)$.}
\end{center}
\end{figure}

\section{Acknowledgments}

This work was partly supported by Grant-in-Aid for Scientific Research (C) (Grant No. 17K05520) from MEXT of Japan and by Nihon University College of Science and Technology Grant-in-Aid for Research. Materials synthesis at Princeton was supported as part of the Institute for Quantum Matter, an Energy Frontier Research Center funded by the U.S. Department of Energy, Office of Science, Basic Energy Sciences under Award No. DE-SC0019331.

\section{APPENDIX: RESULTS OF $C_{11}(T)$}

Figures 5(a) and 5(b) respectively depict the compression modulus $C_{11}(T)$ of NaCaCo$_2$F$_7$ with $H=0$ below 200 K and that with $H||$[110] below 20 K. As seen in Fig. 5(a), $C_{11}(T)$ exhibits monotonic hardening upon cooling down to $\sim$20 K, as is usually observed in solids [\cite{Varshni}]. However, at low temperatures below $\sim$20 K, $C_{11}(T)$ exhibits unusual softening upon cooling [the inset in Fig. 5(a)]. As seen in Fig. 5(b), this softening is suppressed by the application of $H$. At temperatures above $\sim$20 K, $C_{11}(T)$ is independent of magnetic field (not shown).


\begin{thebibliography}{prb}
\bibitem{Ramirez}	A. P. Ramirez, Annu. Rev. Mater. Sci. {\bf 24}, 453 (1994).
\bibitem{Lee0}	S. -H. Lee, H. Takagi, D. Louca, M. Matsuda, S. Ji, H. Ueda, Y. Ueda, T. Katsufuji, J. -H. Chung, S. Park, S. -W. Cheong, and C. Broholm, J. Phys. Soc. Jpn. {\bf 79}, 011004 (2010).
\bibitem{Gardner}	J. S. Gardner, M. J. P. Gingras, and J. E. Greedan, Rev. Mod. Phys. {\bf 82}, 53 (2010).
\bibitem{Krizan1}	J. W. Krizan and R. J. Cava, Phys. Rev. B {\bf 89}, 214401 (2014).
\bibitem{Krizan2}	J. W. Krizan and R. J. Cava, J. Phys.: Condens. Matter {\bf 27}, 296002 (2015).
\bibitem{Krizan3}	J. W. Krizan and R. J. Cava, Phys. Rev. B {\bf 92}, 014406 (2015).
\bibitem{Sanders}	M. B. Sanders, J. W. Krizan, K. W. Plumb, T. M. McQueen, and R. J. Cava, J. Phys.: Condens. Matter {\bf 29}, 045801 (2017).
\bibitem{Ratcliff}	W. Ratcliff II, S.-H. Lee, C. Broholm, S.-W. Cheong, and Q. Huang, Phys. Rev. B {\bf 65}, 220406(R) (2002).
\bibitem{Saunders}	T. E. Saunders and J. T. Chalker, Phys. Rev. Lett. {\bf 98}, 157201 (2007).
\bibitem{Andreanov}	A. Andreanov, J. T. Chalker, T. E. Saunders, and D. Sherrington, Phys. Rev. B {\bf 81}, 014406 (2010).
\bibitem{Shinaoka}	H. Shinaoka, Y. Tomita, and Y. Motome, Phys. Rev. Lett. {\bf 107}, 047204 (2011).
\bibitem{Silverstein}	H. J. Silverstein, K. Fritsch, F. Flicker, A. M. Hallas, J. S. Gardner, Y. Qiu, G. Ehlers, A. T. Savici, Z. Yamani, K. A. Ross, B. D. Gaulin, M. J. P. Gingras, J. A. M. Paddison, K. Foyevtsova, R. Valenti, F. Hawthorne, C. R. Wiebe, and H. D. Zhou, Phys. Rev. B {\bf 89}, 054433 (2014).
\bibitem{Ross2} K. A. Ross, J. M. Brown, R. J. Cava, J. W. Krizan, S. E. Nagler, J. A. Rodriguez-Rivera, and M. B. Stone, Phys. Rev. B {\bf 95}, 144414 (2017).
\bibitem{Ross1} K. A. Ross, J. W. Krizan, J. A. Rodriguez-Rivera, R. J. Cava, and C. L. Broholm, Phys. Rev. B {\bf 93}, 014433 (2016).
\bibitem{Frandsen} B. A. Frandsen, K. A. Ross, J. W. Krizan, G. J. Nilsen, A. R. Wildes, R. J. Cava, R. J. Birgeneau, and S. J. L. Billinge, Phys. Rev. Mat. {\bf 1}, 074412 (2017).
\bibitem{Sarkar} R. Sarkar, J. W. Krizan, F. Br\"uckner, E. C. Andrade, S. Rachel, M. Vojta, R. J. Cava, and H.-H. Klauss, Phys. Rev. B {\bf 96}, 235117 (2017).
\bibitem{Zeisner} J. Zeisner, S. A. Br\"auninger, L. Opherden, R. Sarkar, D. I. Gorbunov, J. W. Krizan, T. Herrmannsd\"orfer, R. J. Cava, J. Wosnitza, B. B\"uchner, H.-H. Klauss, and V. Kataev, Phys. Rev. B {\bf 99}, 155104 (2019).
\bibitem{Luthi} B. L\"uthi, {\it Physical Acoustics in the Solid State} (Springer, 2005).
\bibitem{Watanabe1}	T. Watanabe, S. Hara, and S. Ikeda, Phys. Rev. B {\bf 78}, 094420 (2008).
\bibitem{Watanabe2}	T. Watanabe, S. Hara, S. Ikeda, and K. Tomiyasu, Phys. Rev. B {\bf 84}, 020409(R) (2011).
\bibitem{Watanabe3}	T. Watanabe, S. Ishikawa, H. Suzuki, Y. Kousaka, and K. Tomiyasu, Phys. Rev. B {\bf 86}, 144413 (2012).
\bibitem{Nii} Y. Nii, N. Abe, and T. Arima, Phys. Rev. B {\bf 87}, 085111 (2013). 
\bibitem{Watanabe4}	T. Watanabe, T. Ishikawa, S. Hara, A. T. M. N. Islam, E. M. Wheeler, and B. Lake, Phys. Rev. B {\bf 90}, 100407(R) (2014).
\bibitem{Watanabe5}	T. Watanabe, S. Takita, K. Tomiyasu, and K. Kamazawa, Phys. Rev. B {\bf 92}, 174420 (2015).
\bibitem{Watanabe6}	T. Watanabe, S. Yamada, R. Koborinai, and T. Katsufuji, Phys. Rev. B {\bf 96}, 014422 (2017).
\bibitem{Watanabe7}	T. Watanabe, S. Kobayashi, Y. Hara, J. Xu, B. Lake, J.-Q. Yan, A. Niazi, and D. C. Johnston, Phys. Rev. B {\bf 98}, 094427 (2018).
\bibitem{Varshni} Y. P. Varshni, Phys. Rev. B {\bf 2}, 3952 (1970).
\bibitem{Kino} Y. Kino, L\"uthi, and M. E. Mullen, J. Phys. Soc. Jpn. {\bf 33}, 687 (1972); Solid State Commun. {\bf 12}, 275 (1973).
\bibitem{Kataoka} M. Kataoka and J. Kanamori, J. Phys. Soc. Jpn. {\bf 32}, 113 (1972).
\bibitem{Hazama} H. Hazama, T. Goto, Y. Nemoto, Y. Tomioka, A. Asamitsu, and
Y. Tokura, Phys. Rev. B {\bf 62}, 15012 (2000).
\bibitem{Yamashita} Y. Yamashita and K. Ueda, Phys. Rev. Lett. {\bf 85}, 4960 (2000).
\bibitem{Tchernyshyov} O. Tchernyshyov, R. Moessner, and S. L. Sondhi, Phys. Rev. Lett. {\bf 88}, 067203 (2002).
\bibitem{Ji} S. Ji, S. -H. Lee, C. Broholm, T. Y. Koo, W. Ratcliff, S. -W. Cheong, and P. Zschack, Phys. Rev. Lett. {\bf 103}, 037201 (2009).
\bibitem{Lee1}	S. -H. Lee, C. Broholm, W. Ratcliff II, G. Gasparovic, Q. Huang, T. H. Kim, and S. W. Cheong, Nature (London) {\bf 418}, 856 (2002).
\bibitem{Tomiyasu1}	K. Tomiyasu, H. Suzuki, M. Toki, S. Itoh, M. Matsuura, N. Aso, and K. Yamada, Phys. Rev. Lett. {\bf 101}, 177401 (2008).
\bibitem{Tomiyasu2}	K. Tomiyasu, T. Yokobori, Y. Kousaka, R. I. Bewley, T. Guidi, T. Watanabe, J. Akimitsu, and K. Yamada, Phys. Rev. Lett. {\bf 110}, 077205 (2013).
\bibitem{Tomiyasu3} K. Tomiyasu, H. Ueda, M. Matsuda, M. Yokoyama, K. Iwasa, and K. Yamada, Phys. Rev. B {\bf 84}, 035115 (2011).
\bibitem{Tomiyasu4}	K. Tomiyasu and K. Kamazawa, J. Phys. Soc. Jpn. {\bf 80}, SB024 (2011).
\bibitem{Tomiyasu5} K. Tomiyasu, M. K. Crawford, D. T. Adroja, P. Manuel, A. Tominaga, S. Hara, H. Sato, T. Watanabe, S. I. Ikeda,J. W. Lynn, K. Iwasa, and K. Yamada, Phys. Rev. B {\bf 84}, 054405 (2011).
\bibitem{Tomiyasu6} K. Tomiyasu, K. Iwasa, H. Ueda, S. Niitaka, H. Takagi, S. Ohira-Kawamura, T. Kikuchi, Y. Inamura, K. Nakajima, and K. Yamada, Phys. Rev. Lett. {\bf 113}, 236402 (2014).
\bibitem{Plumb} K. W. Plumb, H. J. Changlani, A. Scheie, S. Zhang, J. W. Krizan, J. A. Rodriguez-Rivera, Y. Qiu, B. Winn, R. J. Cava, and C. L. Broholm, Nature Phys. {\bf 15}, 54 (2019).
\bibitem{Zhang}	S. Zhang, H. J. Changlani, K. W. Plumb, O. Tchernyshyov, and R. Moessner, Phys. Rev. Lett. {\bf 122}, 167203 (2019).
\end{thebibliography}
\end{document}